# THERMOELECTRIC PHENOMENON IN HOLLOW BLOCKS

M. Wehbe[1], J. Dgheim[1]*, E. Sassine[1]

[1]Laboratory of Applied Physics (LPA), Group of Mechanical, Thermal & Renewable Energies (GMTER), Lebanese University, Faculty of Sciences II.

*Corresponding Author Email: jdgheim@ul.edu.lb

## ABSTRACT

The work presented in this article describes thermoelectric effect in hollow blocks for heat waste harvesting purposes. The study consists of developing a numerical model formed by a heat transfer equation coupled to thermoelectric effects equations to study thermoelectric generators(TEG) incorporated inside Lebanese hollow blocks through two simulations using finite difference scheme and using finite element scheme. Results showed a voltage of 5.85mV produced from a single 8.6 x 0.4 x 0.4 cm$^3$ thermoelectric leg made of Bismuth Antimony Telluride for $\Delta T$=30K. A design with 3 TEGs incorporated inside a hollow block was tested and validated numerically using both methods, the main results obtained for $\Delta T$=30K, showed a voltage $\Delta V$=0.72V, a current $I$=0.06 A and a figure of merit $ZT$=0.55. The design was then optimized for economic purposes.

**Nomenclature**

*T*: Temperature [K]  
*u*, *v*: Velocity [m/s]  
*ρ:* Density [Kg/m$^3$]  
*V*: Voltage [V]  
*S:* Surface [m$^2$]  
*I*: Electric current [A]  
*τ*: Thomson coefficient [V/K]  
*α*: Seebeck coefficient [V/K]  
**m:** Mass [Kg]  
*K*: Thermal conductivity [W/m K]  
*r*: Electric resistivity [Ω.m]  
*ΔT*: Temperature difference [K]  
*ɳ*: Efficiency  
*R:* Resistor [Ω]  

Δ*V:* Electric potential [V]  
*σ*: Electric conductivity [S/m]  
*ZT*: Merit factor  
*P:* Power [W]  
*E*: Electric field density [V/m]  
*Q:* Heat energy [J.s$^{-1}$]  
*J*: Current density vector [A.m$^2$]  
*q*: Thermal power [W]  
*C$_p$*: Specific heat for constant pressure [J/Kg. K]  
*π*: Peltier coefficient [V]  
*∇T*: Temperature gradient [K.m$^{-1}$]  
**t:** Time [s]  
*ɳ$_{reduced}$:* Reduced efficiency  

## I. Introduction

The international interest linked to energy recovery continues to grow in order to limit the use of fossil fuels and the consumption of natural resources. Energy recovery is particularly interesting when there is a correspondence between the quantity of recoverable energy and the quantity of energy required for an application. This recovered energy can come from different sources such as temperature gradient (thermoelectric effect). Thermoelectricity was discovered around 200 years ago, it's a transformation of heat into electricity in materials (Seebeck effect) and conversely of electricity into heat or cold (Peltier effect). Waste energy can be harvested using thermoelectric generator(TEG). Thermoelectric generator are devices constructed of two essential junctions p-type and n-type able to generate a voltage when a temperature gradient is applied. The performance of TEG is characterized by the merit factor *ZT* and given by $ZT = \alpha^2 \frac{\sigma T}{K}$. Good thermoelectric materials have large coefficient *α*, with low thermal conductivity *K* to maintain a temperature gradient, and a high electric conductivity.

A historic sector of thermoelectric generators is the space exploration where a radioisotope thermoelectric generator (RTG) [1] that consists of a thermoelectric generator with a nuclear-powered thermal generator as a backup converts heat coming from radioactive atoms emissions into electricity. The automotive sector, where the market for cleaner cars is fierce and government support is strong, is the most active field for energy recovery: Orr et al. [2] recycled waste heat from car engine using 8 thermoelectric modules (62mmx62 mm each). The maximum power produced was 38W, with a 2.46% efficiency for *ΔT* =105 ºC [3]. TEG has also application in the micro production for sensors and microelectronics domain: devices such as MPG-D751 module with dimensions of 4.2mm x 3.3mm x 1mm, produces minimum 1mW for a temperature difference of 10 ° C, and 60 mW of electric power for temperature difference of 60 K, which is the power needed for most micro sensors[4].

The building material industry is having an international interest related to the energy efficiency with the purpose to build new sustainable construction devices. One of TEG's applications is power generation from building walls; with good insulation, a temperature difference between the outside surface and inside surface of walls can be enough to create electricity. The electric voltage generated can be sufficient to power smart sensors and devices which can help in cutting energy consumption and leading to more passive houses, therefore maintaining a comfortable climate. Byon et al. [5] used the HMN-6055 TEG in both summer and winter, for $\Delta T$ = 50 ° C, average voltage produced was 0.3 V, and average power was 0.01 W and 0.03 W in the extreme days. The average amount of electric energy produced per day was about 0.1 Wh. Three to four energy-harvesting blocks were shown to be capable of supplying enough electricity for sensors [6].

Sassine et al. [7] studied Lebanese hollow block for energy harvesting purposes. They noticed through numerical studies that the addition of a row of cavities creates more separation in the temperature distribution inside the blocks keeping one part warm and the other part cold while adding adjacent cavities reduces slightly the air velocity in the cavities which will improve the thermal comfort inside the building and reduce thermal losses across the wall's surfaces[7]. They also found experimentally and numerically that expanded polystyrene (EPS) Concrete Hollow Block have a high thermal resistance of 0.31 $m^2$ K $W^{-1}$ and can reach higher values [8] thus a low thermal conductivity. Therefore, in this work thermoelectric effect will be studied on this type of Lebanese block of parpaing with expanded polystyrene (EPS) Concrete.

Bismuth telluride ($Bi_2Te_3$) occupies a prominent position among thermoelectric materials due to its high efficiency in converting any form of waste heat into renewable electrical energy. For doping modification, Sb doping leaves a very promising impact on $Bi_2Te_3$-based alloy's thermoelectric properties [9]. In our work bismuth antimony telluride ($Bi_{0.5}Sb_{0.5})_2Te_3$ is used after alloying $Bi_2Te_3$ with $Sb_2Te_3$. With the creation of ternary solid solutions (Bi, Sb, Te) we are able to control the concentration of carriers and achieve the corresponding figure of merit value.

**II. Mathematical Model**

The physical model is formed by a two-dimensional cell (Figure 1(a)) filled with a thermoelectric material. Two sources of heat are placed on both sides of the cell. They are used to adjust the temperatures of the two opposite sides.

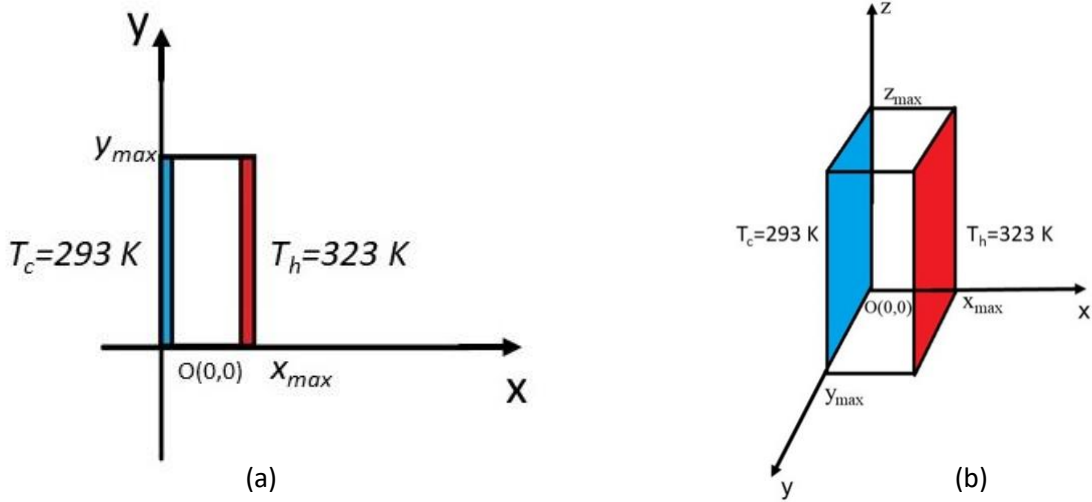

**Figure 1:** (a) Physical model of 2D cell and (b) physical model of 3D cell.

The hot temperature $T_h$ and the cold temperature $T_c$ shown in figure 1(a) are fixed to certain values, considering that the other sides of the cell which are parallel to the *x*-axis and *y*-axis are adiabatic. O (0,0) is considered the origin of the system. During operation, the module receives a certain amount of heat from the hot source. The thermoelectric module converts this thermal energy into work in the form of an electric output current *I*.

To simplify our mathematical model, some conditions are imposed:
- Mathematical equations are transient and two-dimensional.
- The surface is considered adiabatic (the heat cannot be transferred outside the system) ($Q = \frac{dT}{dx} = 0$ for $x = 0$ and $x = x_{max}$ )
- The pressure is the atmospheric pressure.
- The velocities are equal to zero in order to describe a solid. (*U,V*=0)

Using the simplifying conditions, the thermoelectric model is explained below:

$$\rho Cp \frac{\partial T}{\partial t} = \nabla[K \nabla T] + \frac{J^2}{\sigma_T} - \tau J \nabla T \quad (1)$$

$\frac{J^2}{\sigma_T}$ is the term representing the energy variation due to joule and Thomson effect.

The term of the Peltier-Thomson effect is written as: $TJ\nabla\alpha$

This term contains both Thomson's and Peltier's contributions.

From Kelvin's relation: $\pi = \alpha T$, the term of the Peltier-Thomson effect is represented as follows

$$TJ\nabla\alpha = TJ\nabla\frac{\pi}{T} = TJ\left[-\frac{1}{T}\nabla\pi - \frac{1}{T^2}\pi\nabla T\right] = J[\nabla\pi - \alpha\nabla T] \quad (2)$$

In the case of a pure Peltier effect $\nabla T=0$, then: $J [\nabla\pi -\alpha\nabla T] = J\nabla\pi$ (3)
In the case of a pure Thomson effect: $\alpha = f(T)$

Then: $J [\nabla\pi -\alpha\nabla T] = J [\frac{d\pi}{dT} - \alpha] \nabla T = \tau J \nabla T$ (4)

with: $\tau = \frac{d\pi}{dT} -$

The figure of merit ZT is given by:

$$ZT = \alpha^2 \frac{\sigma T}{K} \tag{5}$$

The generated power is given by:

$$P = \eta q h = \frac{T_h - T_c}{T_h} \frac{\sqrt{1+ZT}-1}{\sqrt{1+ZT}+\frac{T_c}{T_h}} \, qh = \alpha \Delta T I - I^2 R \tag{6}$$

Where $\eta_e = \frac{T_h - T_c}{T_h}$ is Carnot efficiency  (7)

And $\eta_{reduced} = \frac{\sqrt{1+ZT}-1}{\sqrt{1+ZT}+\frac{T_c}{T_h}}$ is the reduced efficiency  (8)

The Seebeck coefficient and electrical conductivity are represented by the following equations deduced from [12].

- $\alpha = -1.111 T^3 \times 10^{-12} - 1.543 T^2 \times 10^{-9} + 1.483 T \times 10^{-6} - 5.729 \times 10^{-5}$
- $\sigma = 0.002667 T^3 + 0.4143 T^2 - 1153 T + 2.943 \times 10^5$

In order to complete our mathematical model, initial and boundary conditions have been added:

- Initial conditions ($t < t_0$) : $T_0 = 293.15$ K $\forall \, x, y$
- Boundary conditions ($t > t_0$) : $x = 0$ ; $T = T_c = 293$ K $\forall \, y$ & $x = x_{max}$ ; $T = T_h = 323$ K $\forall \, y$

$$y = 0 \; ; \; Q = \frac{dT}{dx} \Big)_{x=0} = 0 \; \forall x \; \text{(adiabatic surface)}$$

$$y = y_{max} \; ; \; Q = \frac{dT}{dx} \Big)_{xmax=0} = 0 \; \forall x \; \text{(adiabatic surface)}$$

## III. Numerical Techniques

Mathematical modeling provides a description of various relations between quantities in a particular circumstance. This can lead to a one-of-a-kind solution that can only be found numerically.

### III.1 Explicit Finite Difference Scheme

The discretized heat equation will be as follows:

$$T^{n+1}(i,j) = [MK^n(i+1,j) - \frac{\Delta t PT^n(i,j)}{\rho Cp \Delta x}] T^n(i+1,j) - [K^n(i,j)(2M+2N)] T^n(i,j) +$$

$$\frac{\Delta t PT^n(i,j)}{\rho Cp} (\frac{1}{\Delta x} + \frac{1}{\Delta y}) T^n(i,j) + MK^n(i-1,j) T^n(i-1,j) + [NK^n(i,j+1) - \frac{\Delta t PT^n(i,j)}{\rho Cp \Delta y}] \, T^n(i,j+1) +$$

$$NK^n(i,j-1) T^n(i,j-1) + \frac{\Delta t J^{2n}(i,j)}{\rho Cp \sigma^n(i,j)}$$

With $M = \frac{\Delta t}{\rho Cp \Delta x^2}$; $N = \frac{\Delta t}{\rho Cp \Delta y^2}$; $PT^n(i,j) = \tau^n(i,j) J^n(i,j)$   (9)

Our approach uses the width $x$ and the length $y$ to describe thermal and electrical transport in a rectangular shape, with $i$ the number of nodes along the $x$-axis and $j$ the number of nodes along the $y$-axis. Finally, the different steps of space ($\Delta x$ and $\Delta y$) and time ($\Delta t$) must be calculated as follows: $\Delta x = \frac{x}{i} = 0.00005$, $\Delta y = \frac{y}{j} = 0.00005$ and $\Delta t = 0.0009$.

## III.2 Finite Element scheme

Dgheim reports that the mathematical model is also solved using the finite element method [10],[11] . As a result, the weak integral of the thermoelectric issue is formulated as follows:

$$w(T, T^*) = \int_V T^* \rho C_p \dot{T} dV - \int_V T^* \nabla.(k\nabla T) dV - \int_V T^* \left(\frac{J^2}{\sigma_T} - \tau J \nabla T\right) dV = 0 \qquad (10)$$

The finite element approximation of equation (10) can be written as:

$$[D]\frac{\{T\}^{n+1}-\{T\}^n}{\Delta t} + ([C] - [H] - [M])\{T\}^n + [L] = 0 \qquad (11)$$

Where, $[D]$ is the storage matrix, $[C]$ is the conductive matrix on a volume, $[H]$ is the conductive matrix in a surface, $[M]$ is the Thomson vector and $[L]$ is the constant matrix.

Equation (11) is solved using an explicit finite difference method and Gauss-Legendre integration.

## IV. Results and Discussion

Implementation of thermoelectric generators inside a hollow block will be presented along with the numerical results of our mathematical model that are validated using the finite difference method (FDM) and the finite element method (FEM).

### IV.1 Model validation

In order to validate our mathematical model, a p-type thermoelectric(TE) element Bismuth Antimony Telluride $(Bi_{0.5}Sb_{0.5})_2Te_3$ is simulated, the results found using the finite element method are compared with the ones found using the finite difference method and the ones found by Martin et al. [12]. After applying $T_c$=273K on the bottom and $T_h$=373K at the top, the variation of the voltage along the length of the leg is shown in Figure 2 for a maximum $\Delta T$=100K.

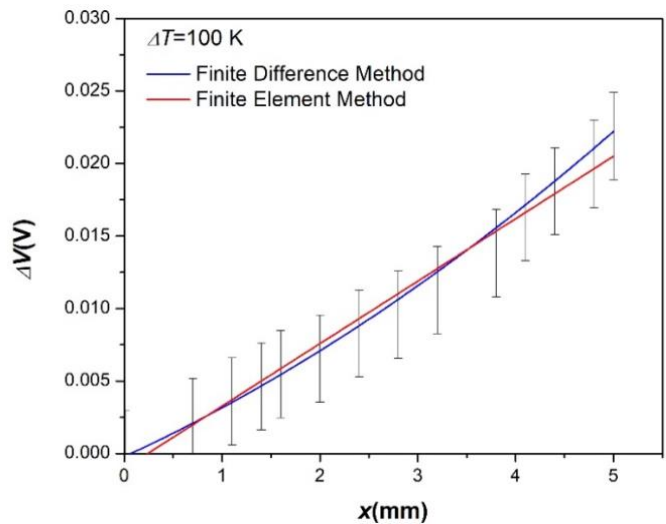

**Figure 2:** Comparison between the results of our model treated by the finite element method and the finite difference method.

The voltage increases with the increase of the sample-length and the temperature differences to reach a maximum value of 20mV with FEM and 21mV with FDM. Both results agree with the one found by Martin et al. [12] with a relative error less than 3%.

Figure 3 shows a comparison between our values and the ones found in literature [12] for the power output as function of current.

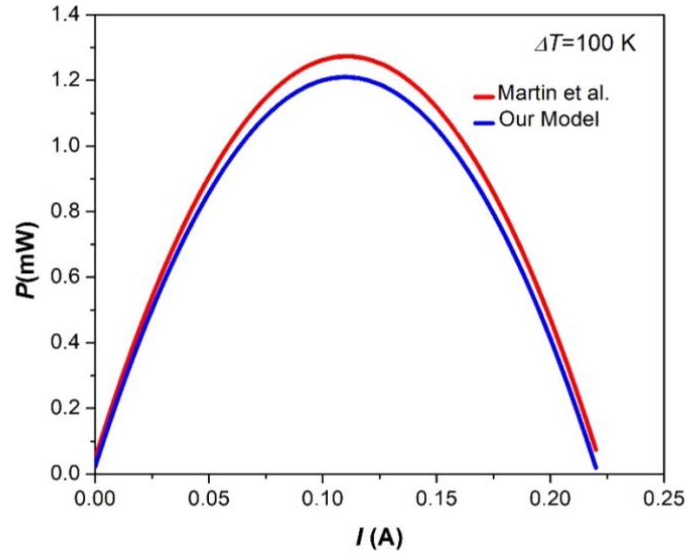

**Figure 3:** Variation of the power output as function of current.

In figure 3 maximum power output is equal to 1.18 mW for $I$=0.108A while with Martin et al. [12] $P_{max}$=1.22mW for $I$= 0.11 A. Relative error is around 3%, therefore our model is validated.

### IV.2 Numerical results

Computational studies are performed using FDM and FEM. Three TEGs are incorporated inside the parpaing block (figure 4) with $T_c$=293 K and $T_h$=323 K; Two lateral with 29 TE elements of $(Bi_{0.5}Sb_{0.5})_2Te_3$ each and one in the bottom with 65 TE elements as shown in figure 4 (b) for $\Delta T$=30 K. The block has size of 10 x 20 x 40 cm$^3$ while one thermoelectric element has a size of 8.6 x 0.4 x 0.4 cm$^3$. Copper electrodes and Aluminum substrates are used to form the TEG.

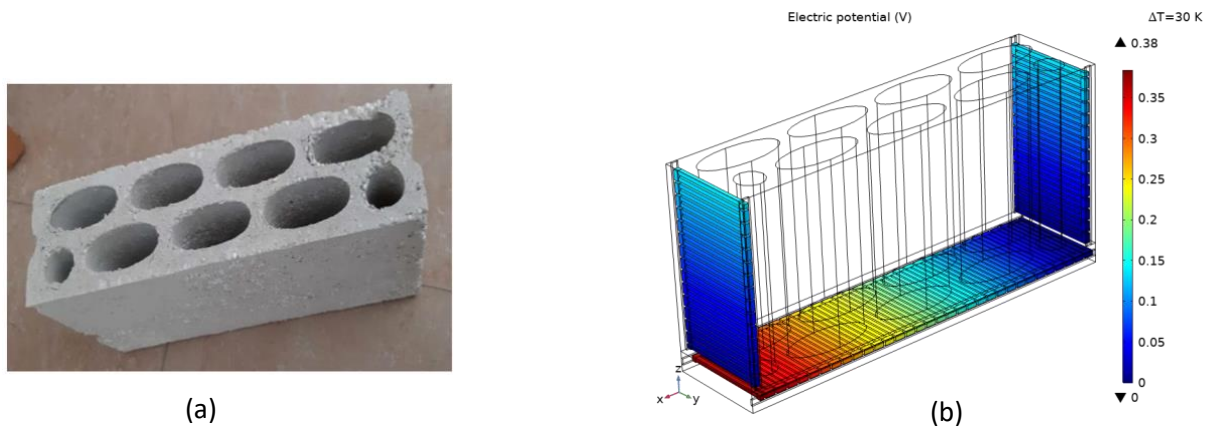

**Figure 4:** (a) Parpaing hollow block [13] and (b) Design of three TEGs inside the block.

Each lateral TEG provides a voltage of 0.17V while a voltage of 0.38V is reached by the bottom TEG for $\Delta T$=30 K. Therefore, with this new design one block of parpaing can provide 0.72 V for $\Delta T$=30 K which means 8.85mV/thermoelectric element. The variation of the voltage for different temperature differences per one block is plotted with two different methods in figure 5.

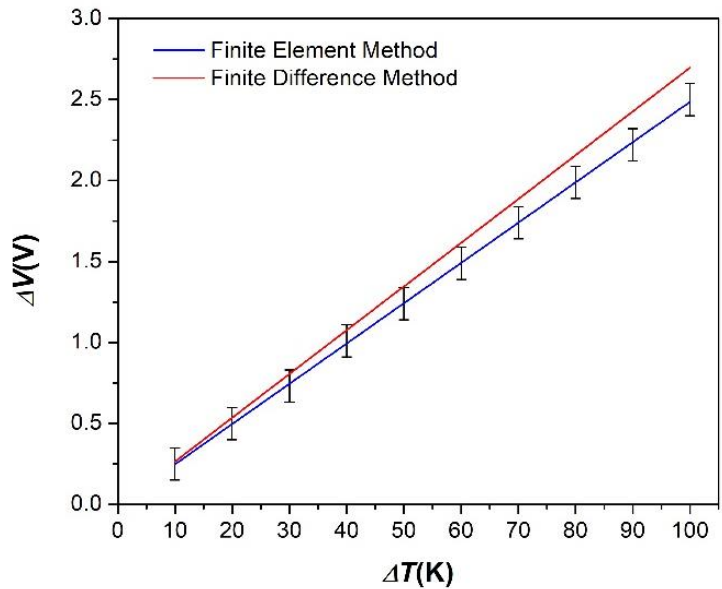

**Figure 5:** Variation of the voltage as function of $\Delta T$ using FDM and FEM.

Table 1 below shows in details the voltage variation and current variation for one block, and an estimation for the number of blocks needed to provide a house with 12 V and 5 A for different $\Delta T$.

| $\Delta T$(K) | 10 | 20 | 30 | 40 | 50 | 60 | 70 | 80 | 90 | 100 |
|---|---|---|---|---|---|---|---|---|---|---|
| $\Delta V$(V) | 0.255 | 0.515 | 0.755 | 1.05 | 1.305 | 1.545 | 1.81 | 2.075 | 2.21 | 2.595 |
| $I$(A) | 0.02 | 0.04 | 0.06 | 0.08 | 0.10 | 0.12 | 0.14 | 0.17 | 0.21 | 0.23 |
| Number of blocks for 12V | 48 | 24 | 16 | 12 | 10 | 8 | 7 | 6 | 6 | 5 |
| Number of blocks for 5 A | 250 | 125 | 84 | 63 | 50 | 42 | 36 | 30 | 24 | 22 |
| Total number of blocks for 12 V and 5 A | 12000 | 3000 | 1344 | 756 | 500 | 336 | 252 | 180 | 144 | 110 |

**Table 1:** Results for different temperature difference.

Power as function of current is presented in figure 6 below.

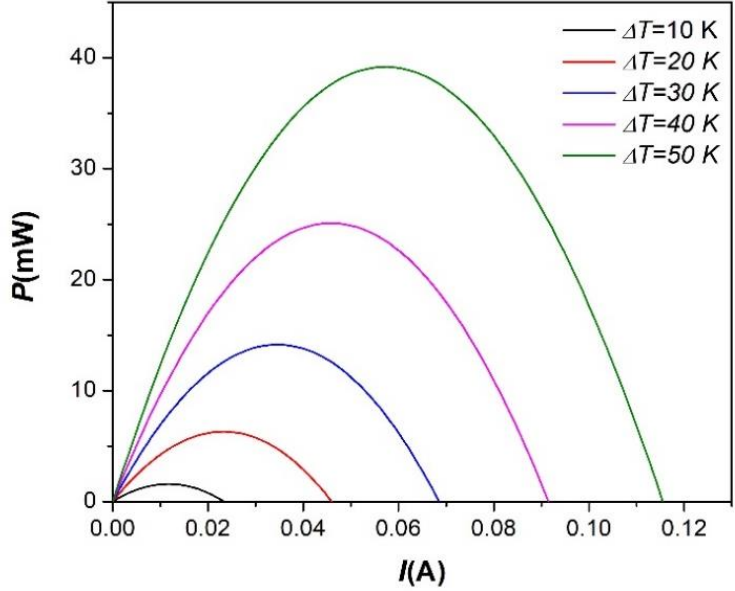

**Figure 6:** Power as a function of current for different $\Delta T$.

For a temperature difference of 30 K, maximum power reached is equal to 14 mW for $I$=0.03A as shown in figure 6. Factor of merit $ZT$ for different $\Delta T$ as function of distance is show in figure 7 (a) and for $\Delta T$=30K (realistic $\Delta T$ for our application), another $ZT$ profile is plotted for different times in figure 7 (b) below where $ZT$ reaches a maximum value of 0.55.

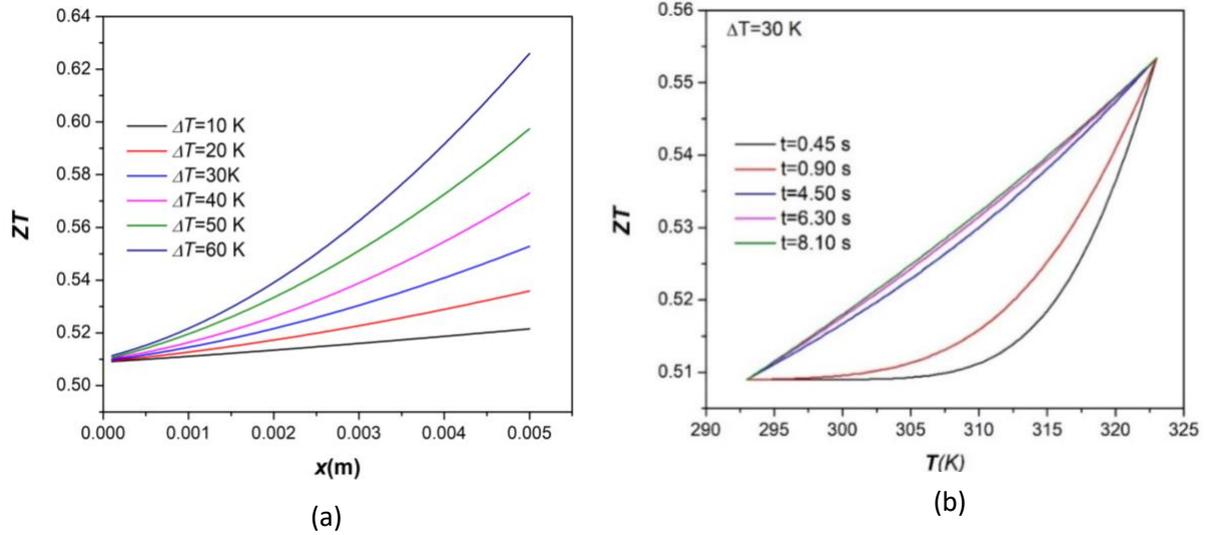

**Figure 7:** (a) $ZT$ as function of distance for different $\Delta T$ and (b) $ZT$ as function of temperature for different times at $\Delta T$=30K.

The efficiency and reduced efficiency are given by:

$\eta_{reduced} = \dfrac{\sqrt{1+ZT}-1}{\sqrt{1+ZT}+\frac{Tc}{Th}}$ and $\eta = \eta_{reduced} * \dfrac{\Delta T}{Th}$ and plotted in figure 8 below at $\Delta T$=30K.

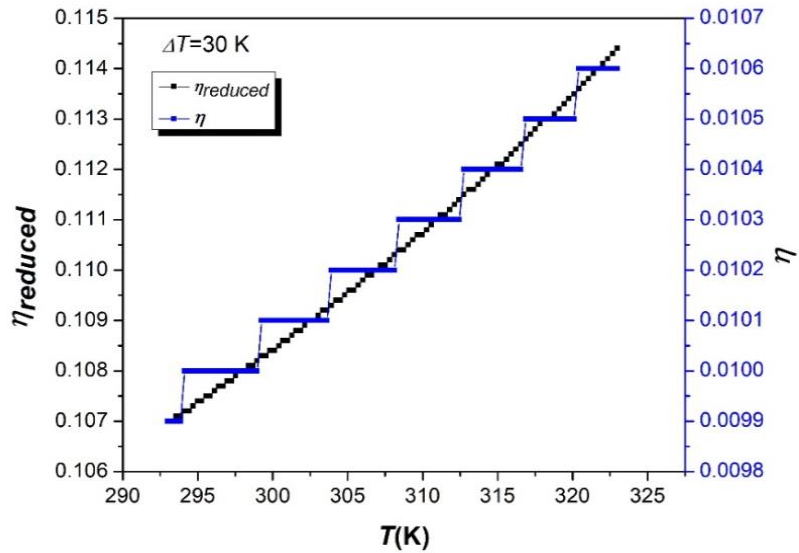

**Figure 8:** Evolution of the efficiency and reduced efficiency as function of Temperature.

The efficiency and reduced efficiency increase with the temperature to reach a maximum value of $1.065 \times 10^{-2}$ and $1.144 \times 10^{-1}$ at $T=323$ K respectively.

## V. Optimization

The cost of this design is one of the criteria affecting its success and ability to be implemented in the real world. The chart below shows the cost of the needed materials.

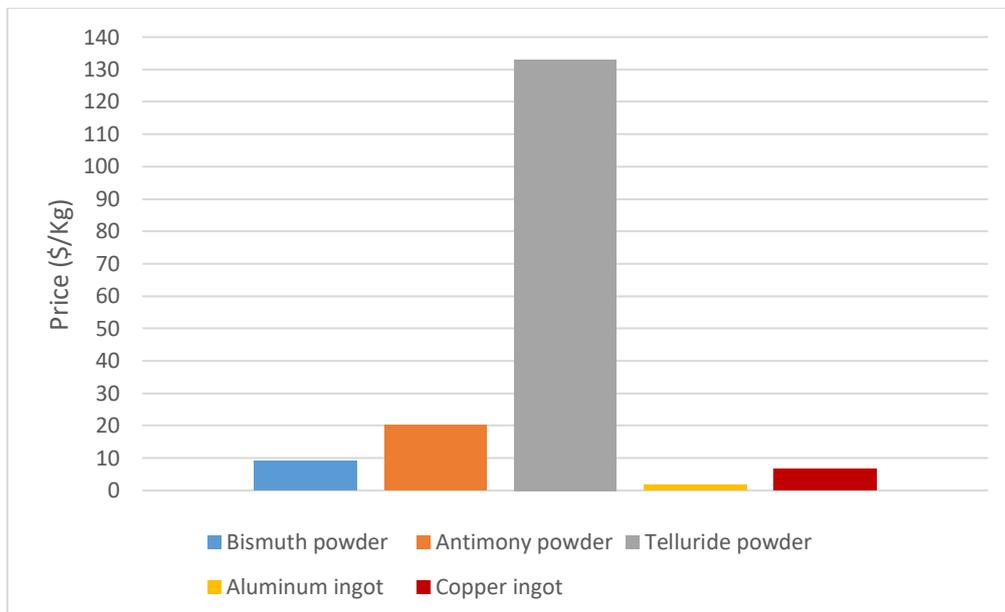

**Figure 9:** Cost of the used materials ($/kg).

One thermoelectric element is made of 53.64% Te in mass, 17.06% Sb in mass and 29.28% Bi in mass. The price of the design presented in figure 4 (b) with a total of 123 thermoelectric elements is approximately equal to 97$. For economic purposes, optimization is performed on

the thermoelectric element's size. These changes in size are chosen in a way to keep a voltage of 8.85mV/thermoelectric element and a current of 0.06 A for $\Delta T$=30K. The length of the copper electrodes and substrates are considered equal for symmetry reasons and the number of TE elements, copper electrodes and substrates is kept the same. We obtained six different models with different thermoelectric leg's dimension presented in table 2.

|  | 1 thermoelectric element | |
| --- | --- | --- |
| Model | Length $L$ (cm) | Thickness $e$ = Width w (cm) |
| 1 | 0.5 | 0.09 |
| 2 | 1 | 0.13 |
| 3 | 1.5 | 0.16 |
| 4 | 2 | 0.19 |
| 5 | 2.5 | 0.21 |
| 6 | 3 | 0.23 |

**Table 2:** Dimensions of six different models.

Simulations of the models are performed using both FDM and FEM to verify the voltage and current results. Figure 10 shows the voltage for different models at $\Delta T$= 30 K with FEM.

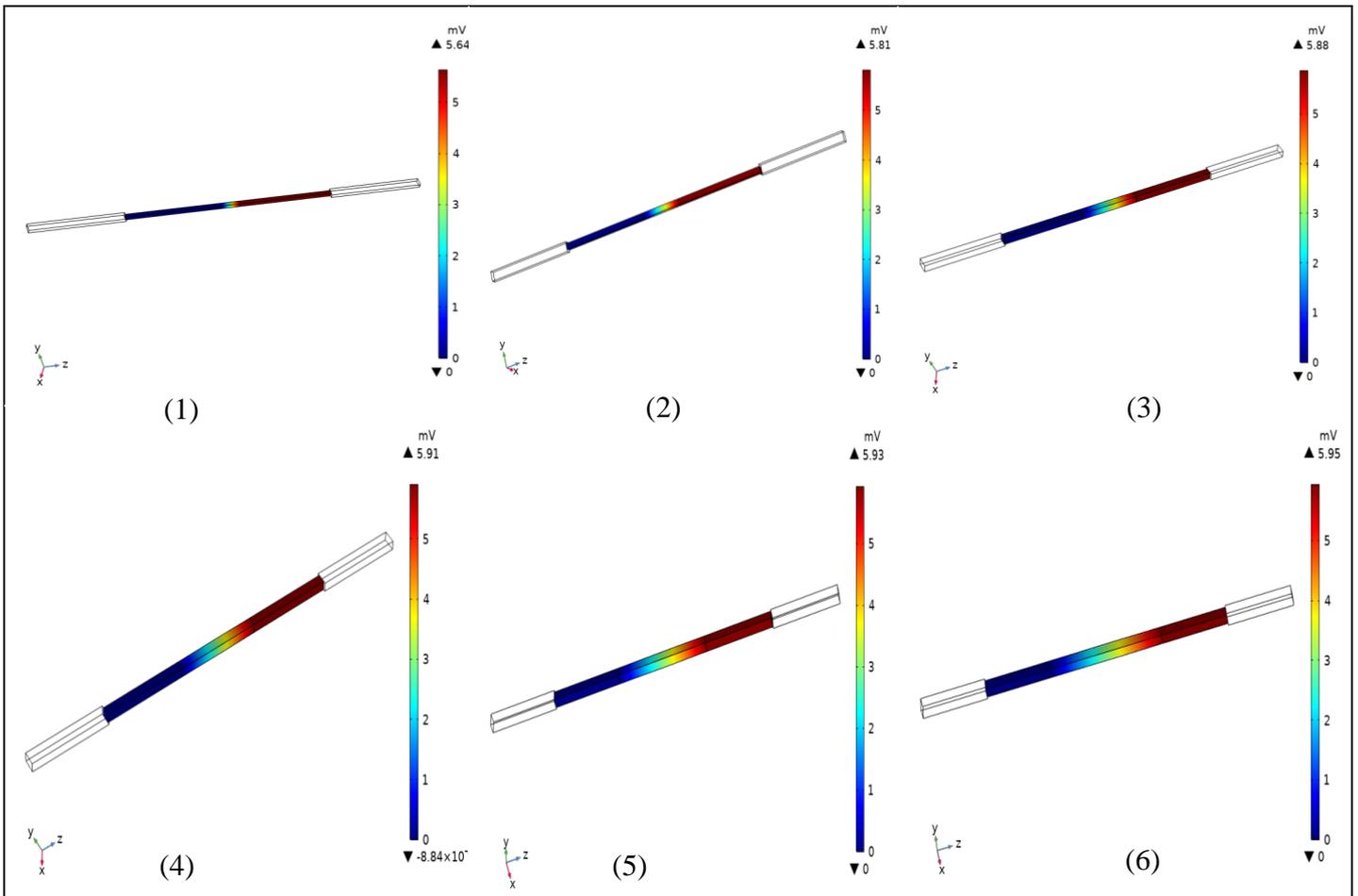

**Figure 10:** Voltage variation for different model at $\Delta T$=30K.

The values of the voltage tested using FEM are approximately equal to 8.85mV. It is clear that for model (1) and (2), $\Delta V$ is less than 8.85mV and this is because the shorter the thermoelectric leg is, the lower is *ΔT* across it. On the other hand, the longer the length of the TE, the higher the voltage (Model 3,4,5...). The current is also validated for the different models using finite difference method as shown in figure 11.

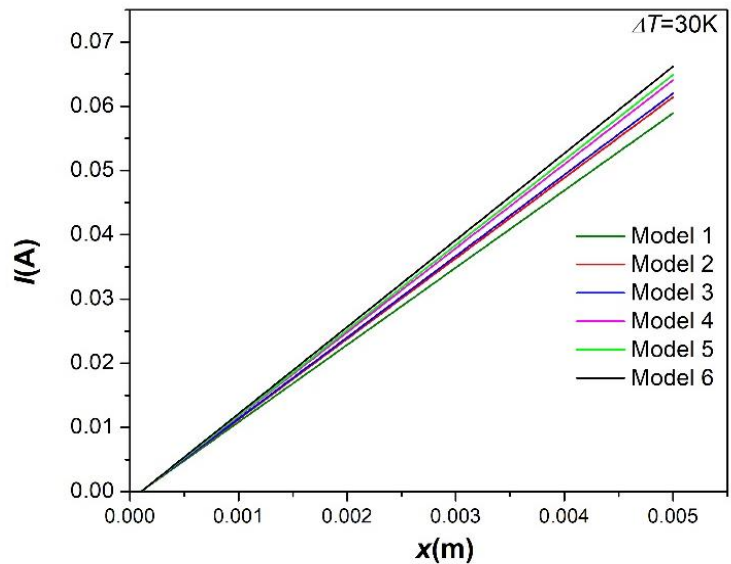

**Figure 11:** Current variation for different Models.

After validation of different TE models for different dimensions, figure 12 below represents the price of one thermoelectric leg for the six models.

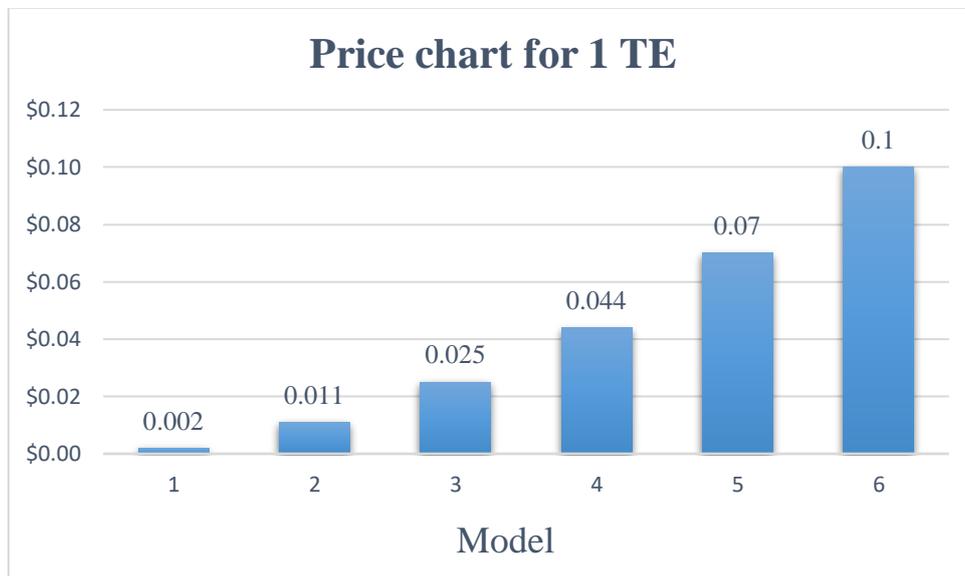

**Figure 12:** Price of one thermoelectric leg for six different Models.

After considering the price of copper electrodes and substrates, the final cost of the 1344 blocks needed to provide a voltage of 12 V and a current of 5 A as shown in table 1 with 123 TE on each block is presented in the chart below for the six different models.

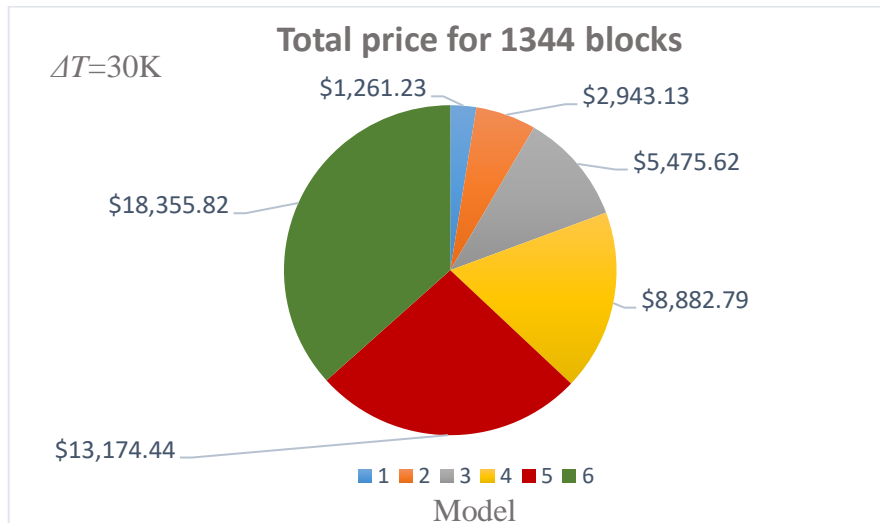

**Figure 13:** Total price to provide 12 V and 5 A for six different Models.

Models with small TE length have an acceptable price, however lowering too much $L$ leads to two inconvenient: 1) $\Delta T$ across the TE will become less than 30K which will reduce the produced voltage and 2) the mechanical strength of the leg might decrease due to the stress caused by high temperature. On the other hand, choosing a model with high TE length results in providing a high voltage but with a very high unaffordable total price.

Therefore, as an optimum model, model (4) with a price around 8880 $ as shown in figure 13 would be a good choice since for $L$=2cm and $e$=0.19 cm, $\Delta V$=5.91mV as shown in figure 10 and $\Delta T$ across the TE's leg becomes equal to 29.5 K only after 30s since aluminum substrates were replaced with copper substrates due to the high thermal conductivity of the latter.

**VI. Conclusion and Perspective**

A mathematical model of heat transfer and thermoelectric effect for a TEG in hollow block is developed using FDM and FEM. Our aim is to utilize heat waste in building walls, therefore, a design of 3 TEGs is created incorporated inside one block of parpaing with a total of 123 TE elements, a voltage of 0.72V is reached, a current $I$=0.06 A and a $ZT$=0.55 for $\Delta T$=30K. This energy harvesting design may be expanded by joining numerous blocks to match the load power's needs. With our design, to provide a house with 12 V and 5 A and for $\Delta T$ of 30K, 1344 blocks are needed. To reduce the implantation cost, different models of thermoelectric element were tested and an optimum one was chosen considering its thermoelectric performance and its price. More research is needed to completely develop the system, including the best installation site for different climates and types of buildings, as well as material selection for more cost-effective manufacturing.

Despite TEG's low efficiency and its high cost, TEGs need very little maintenance and can last up to 20 years with limited performance degradation. Compared to solar photovoltaic, TEGs are independent of external variables and can work not just under solar rays without forgetting that it has a relatively simple design being completely passive with no moving parts.

Many improvements can be performed on our model: 1) As semiconductor technology advances, it is beneficial to study the effect of doping on our material to find the optimal composition of Bismuth Antimony Telluride to enhance the figure of merit and obtain the best thermoelectric performance. 2) A nano-scale performance can also be studied in an attempt to reduce the lattice thermal conductivity significantly while having little influence on electrical conductivity which can be done through engineering lattice with high density of grain boundaries to increase phonons scattering but without affecting the electrons scattering. And finally, we can also benefit from this design in other applications where a higher $\Delta T$ can be reached.